\documentclass[twoside]{ilcws07}
\usepackage[latin1]{inputenc}
\usepackage[dvips]{graphicx,epsfig,color}
\usepackage{wrapfig,rotating}
\usepackage{amssymb,amsmath,array}

\begin{document}
\title{First results of systematic studies done with different types of Silicon Photomultipliers}
\author{
C.~Bosio$^1$, S.~Gentile$^2$, E.~Kuznetsova$^1$, F.~Meddi$^2$
\vspace{.3cm}\\
1- INFN Roma 1\\
Piazzale Aldo Moro 5, 00185 Roma, Italy\\
\vspace{.1cm}\\
2- Universit\`{a} degli Studi di Roma "La Sapienza"\\
Piazzale Aldo Moro 5, 00185 Roma, Italy
}
\maketitle

\begin{abstract}
The presented results are obtained during the
first steps taken in order to develop a setup and measurement procedures
which allow to compare properties of diverse kinds of silicon photomultipliers.
The response to low-intensity light was studied for silicon photomultipliers
produced by CPTA (Russia), Hamamatsu (Japan), ITC-irst (Italy) and SensL (Ireland).
\end{abstract}

\section{Introduction}
Fast development of the silicon photomultiplier technology results in a
number of different sensor types available on market and produced by
various manufacturers. Despite the different commercial names of the
devices, here all of them are referred as SiPM.

The measurements discussed here include current-voltage characteristics
and studies of the SiPM response to low-intensity light. The results of the
former measurements are mostly used to check the sample operability and to
define the range of bias voltages for the latter studies. The measurements
of the SiPM response to the light provide a number of parameters suitable for
the comparison of different samples.

Five samples of silicon photomultipliers produced by the following manufacturers have been studied:
CPTA\footnote{CPTA, Russia, http://www.zao-cpta.ru} 
produced samples distributed by Obninsk University and
Forimtech\footnote{Forimtech SA, http://www.forimtech.ch}, HAMAMATSU
produced Multi-Pixel Photon Counter S10362-11-025C~\cite{hamamatsu}, 
ITC-irst\footnote{ITC-irst, Italy, http://www.itc.it/irst} and 
SensL\footnote{SensL, Ireland, http://www.sensl.com} produced samples 
were compared on the base of the measurement results.

\section{Measurement Setup}
In order to obtain the SiPM response to low-intensity light, 
the sensors were illuminated with a light emitting diode operated in a pulse
mode. The LED drive developed by Institute
of Physics ASCR (Prague)  was used. 
The drive provides current pulses with tunable amplitude and
duration with a sharp rise time down to $2\;\rm{ns}$~\cite{ivo}.
The signal from SiPM was read out with a charge-sensitive preamplifier and digitised with
an integrating ADC. The LED pulse of about $6\;\rm{ns}$ duration and ADC gate of
$65\;\rm{ns}$ width were synchronised by means of a common trigger.

The measurements were done at room temperature.
Temperature variation during the measurements done for one sample
did not exceed  2$^{\rm o}$C, for all measurements discussed here the total variation was
less than  4$^{\rm o}$C.

\section{Measurement Results}
\begin{wrapfigure}{r}{0.5\columnwidth}
\centerline{\includegraphics[width=0.45\columnwidth]{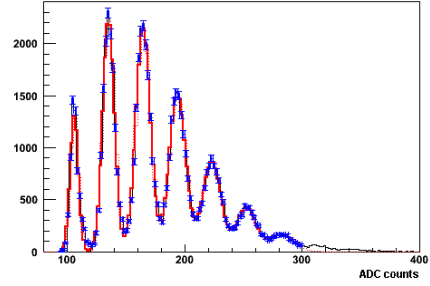}}
\caption{ADC spectrum of the SiPM response to the low-intensity pulsed light
fitted to a superposition of seven gaussian peaks.}\label{fig:spectrum}
\end{wrapfigure}

Fig.~\ref{fig:spectrum} shows an example of the SiPM response to low-intensity
light measured with the described setup. The peaked structure indicates the
number of cells fired during one light pulse, starting with the pedestal for
no cells fired. The distance between peaks corresponds to the SiPM gain.

The spectrum is fitted as a sum of gaussian distributions:
\begin{equation}
 \label{egauss}
 \sum_{i} G(N_i,\mu_i,\sigma_i) =  \sum_{i} G(N_i,\mu_0+i\cdot g,\sigma_i),
\end{equation}
where $\mu_0$ corresponds to the pedestal position and $g$ is gain in units of
ADC counts. From statistical considerations the width of $i^{\rm th}$ peak
$\sigma_i$ can be expressed as 
\begin{equation}
 \label{esigma}
 \sigma_i = \sqrt{\sigma_0^2 + i\cdot\langle\sigma_{px}\rangle^2},
\end{equation}
where $\sigma_0$ is the pedestal width and $\langle\sigma_{px}\rangle$ represents 
fluctuations of the one-cell response averaged over the active area of the sample.

\begin{figure}[bh]
\begin{minipage}{0.45\textwidth}
\includegraphics[width=6.5cm]{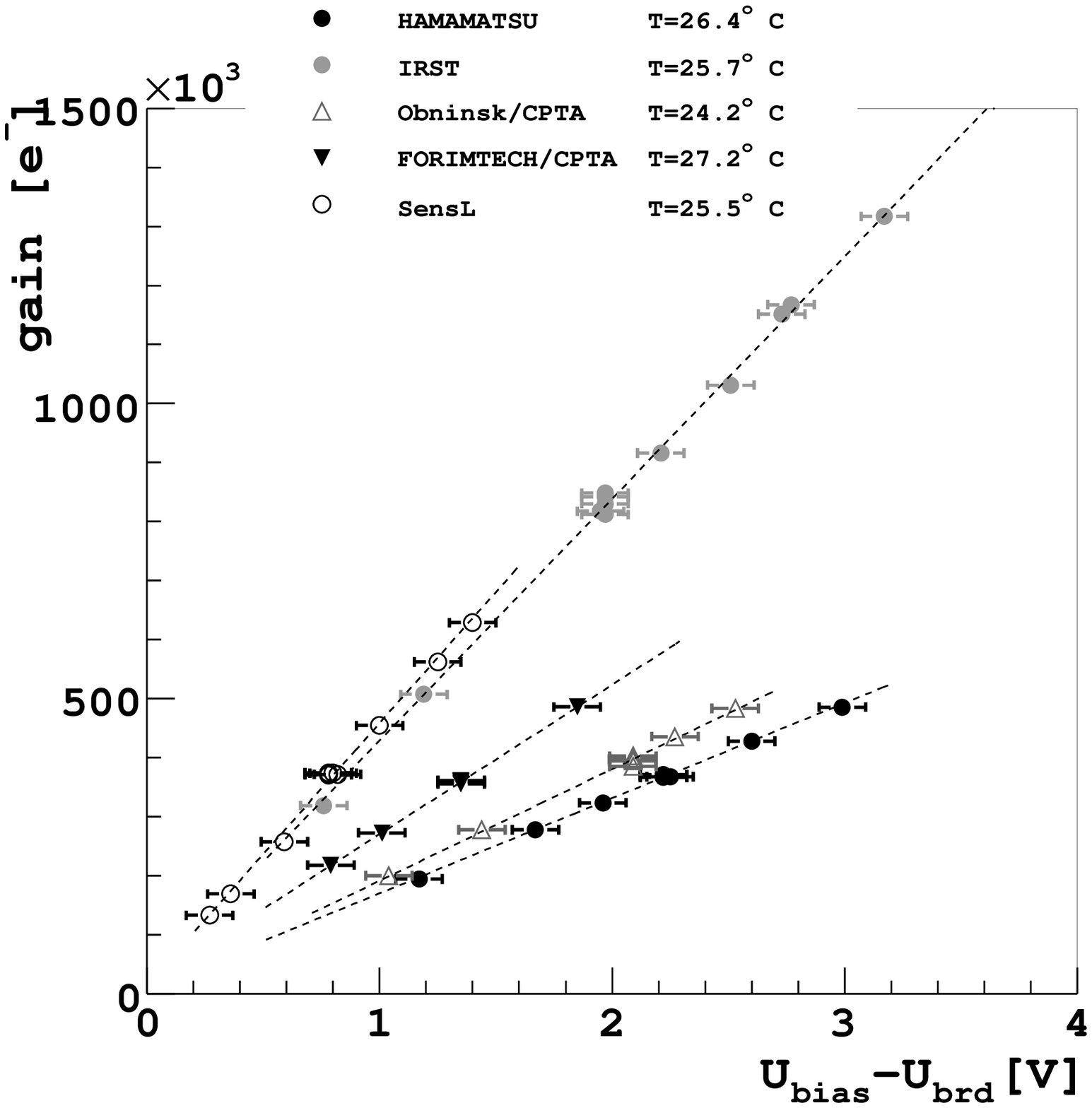}
\end{minipage}
~
\begin{minipage}{0.45\textwidth}
\includegraphics[width=6.5cm]{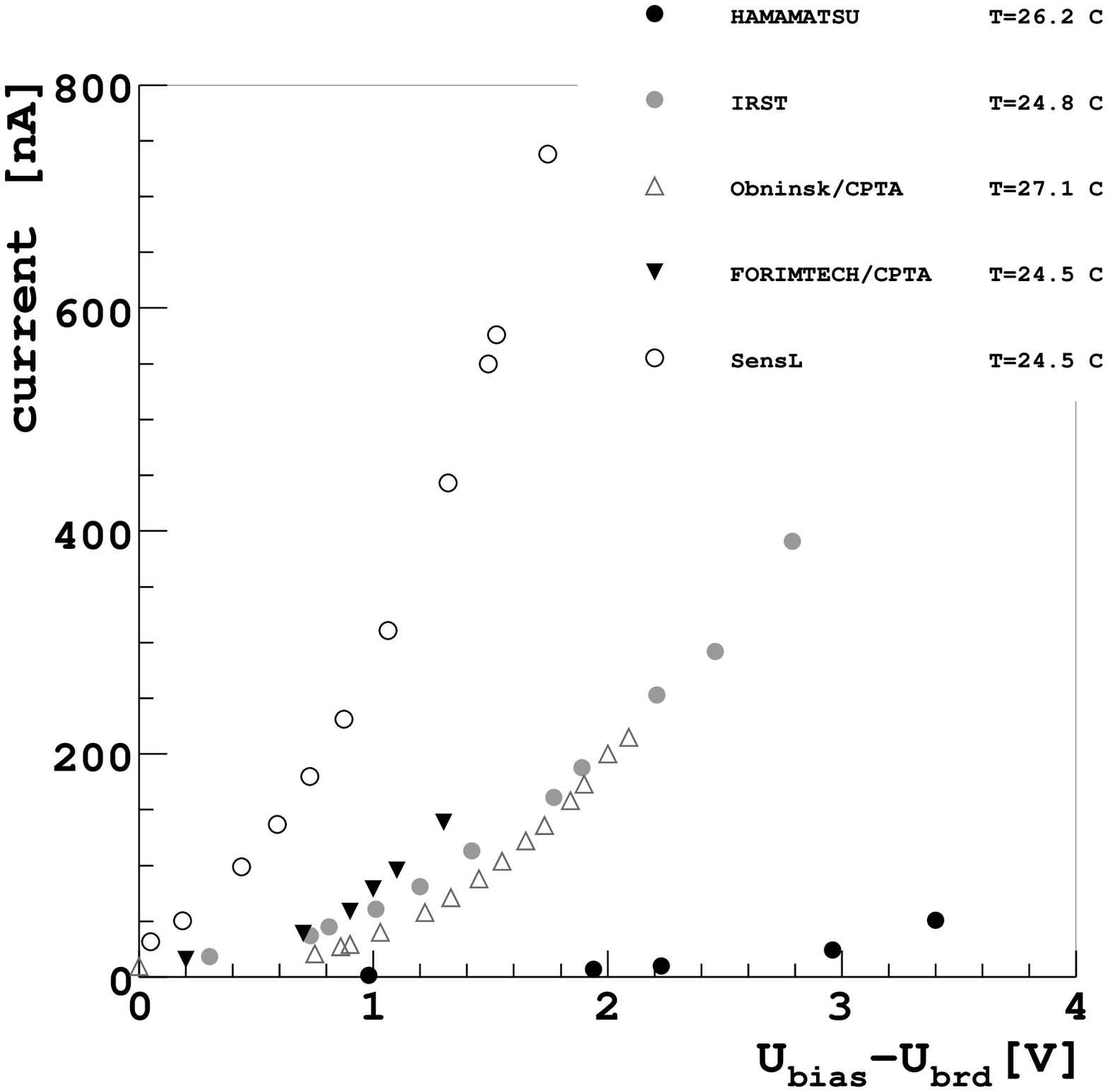}
\end{minipage}
\caption{Gain (left) and dark current (right) as a functions of the overvoltage. }\label{fig:gain}
\end{figure}

The gain as a function of the overvoltage $U_{bias}-U_{brd}$ was studied 
for several values of the bias voltage.
The breakdown voltage $U_{brd}$ is defined here as the bias voltage
corresponding to the gain equal to one.
The results are shown in fig.~\ref{fig:gain} (left). Fig.~\ref{fig:gain} (right) shows 
the dark current as function of the overvoltage.

\begin{figure}[th]
\begin{minipage}{0.45\textwidth}
\includegraphics[width=6.5cm]{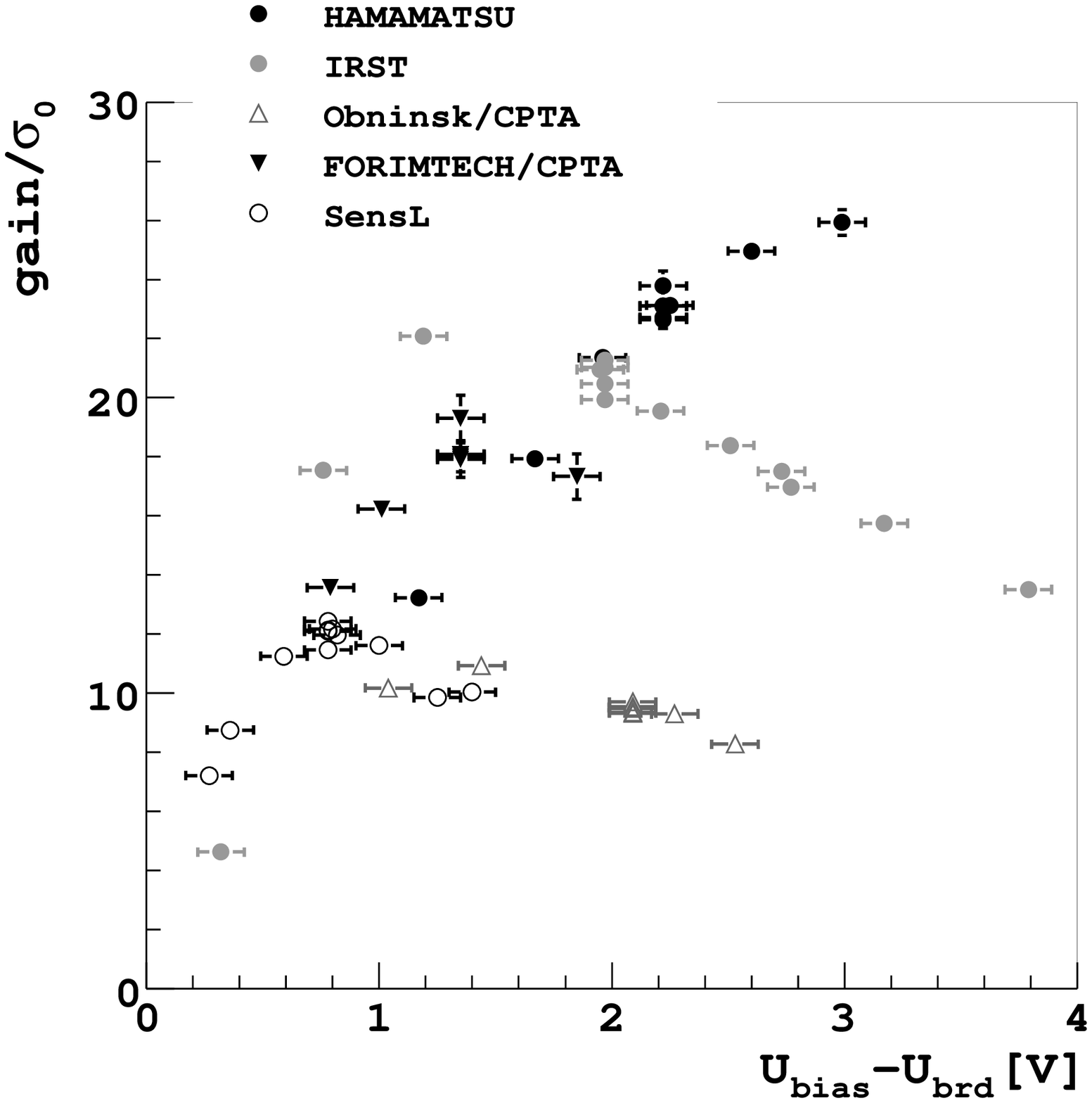}
\end{minipage}
~
\begin{minipage}{0.45\textwidth}
\includegraphics[width=6.5cm]{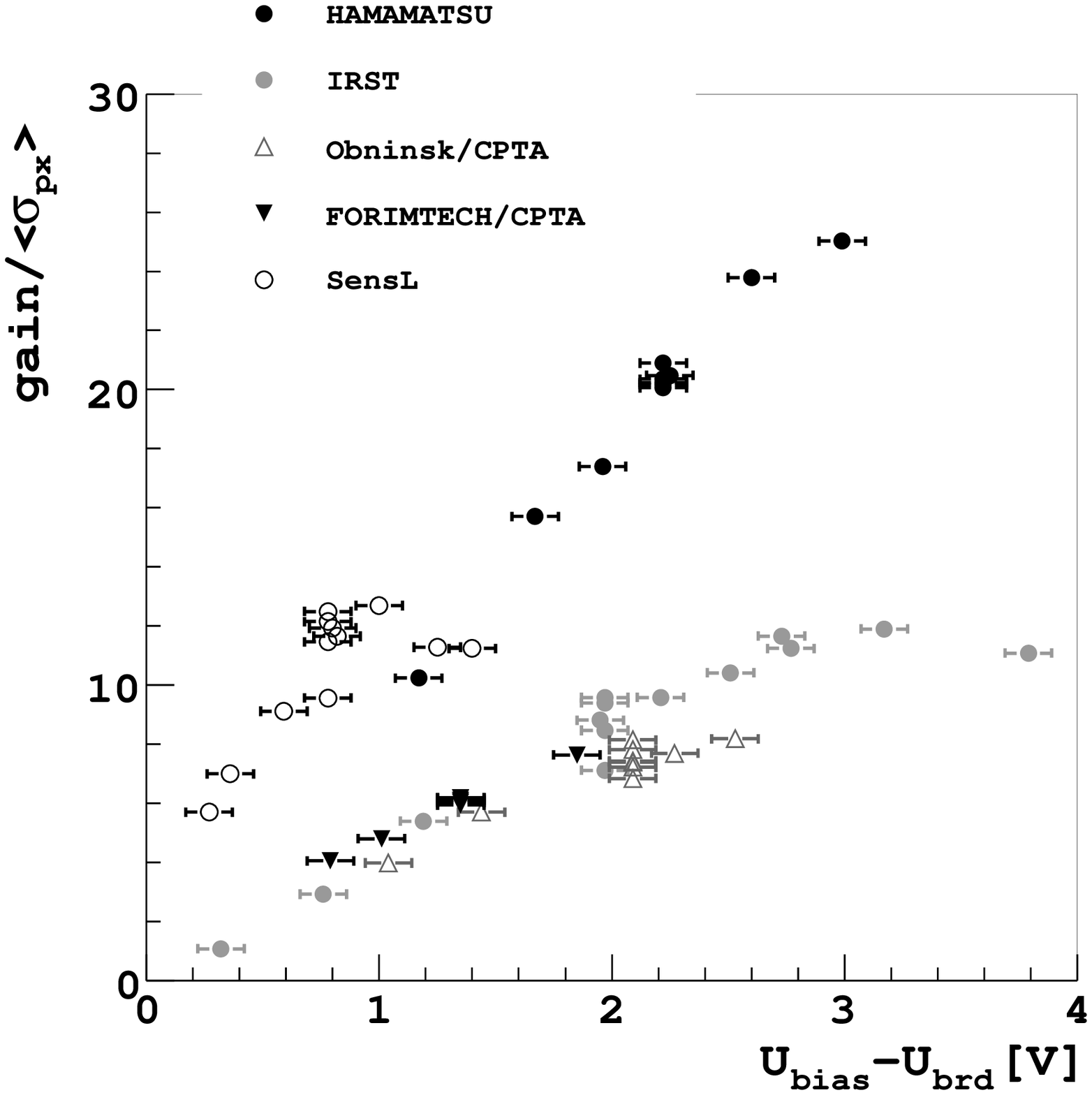}
\end{minipage}
\caption{The gain normalised to the corresponding values of 
pedestal width (left) and $\langle\sigma_{px}\rangle$ (right) as functions
of the overvoltage.}\label{fig:gsigma}
\end{figure}

Fig.~\ref{fig:gsigma} shows gain normalised to the corresponding values of 
pedestal width $\sigma_0$ (left) and $\langle\sigma_{px}\rangle$ (right) as functions
of the overvoltage. Being averaged over the active area of SiPM,
$\langle\sigma_{px}\rangle$ contains statistical and systematic parts:
\begin{equation}
 \label{egsigma}
\langle\sigma_{px}\rangle \sim \sqrt{N + \sigma_{nu}^2},
\end{equation}
where $N$ is the average number of charge carriers in the avalanche and
$\sigma_{nu}$ represents non-uniformity of the amplification over the SiPM
active area. Since gain is proportional to the carriers number,
$g \sim N$, the ratio $g/\langle\sigma_{px}\rangle$
shown in fig.~\ref{fig:gsigma} (right) tends to a linear behaviour for the low
gain values $N<<\sigma_{nu}^2$.
For the high gain values $N>>\sigma_{nu}^2$ the dependence would correspond to
square root low $g/\langle\sigma_{px}\rangle\sim \sqrt{N}$ in the case of a
constant $\sigma_{nu}$. As seen from fig.~\ref{fig:gsigma} (right), the ratio
$g/\langle\sigma_{px}\rangle$
obtained for some of the samples indicates growth of the
non-uniformity factor $\sigma_{nu}$ with the bias voltage.

\section{Conclusion}
The obtained results demonstrated operability and potential of the
developed setup. Further development and tune of the setup, measurement
procedure and data treatment will allow to obtain comparative characteristics
of diverse types of silicon photomultipliers.

\section{Acknowledgements}
We would like to thank 
Prof.~R.~Battiston (University and INFN of Perugia),
Prof.~V.~Saveliev (Obninsk State University and DESY),
Dr.~I.~Polak (Institute of Physics of the ASCR, Prague), Dr.~E.~Popova (MEPhI)
and Mr.~N.~D`Ascenzo (DESY) for their support and fruitful discussions.

%
\begin{footnotesize}
\bibliographystyle{unsrt}
\bibliography{sipm}

\end{footnotesize}


\end{document}